# Amplification of entangled photon states by klystron

Zheng Li[1,2,*]

[1]*Center for Free-Electron Laser Science, DESY,*

*Notkestraße 85, D-22607 Hamburg, Germany*

[2]*SLAC National Accelerator Laboratory, Menlo Park, California 94025, USA*

## Abstract

Entangled photon pairs–discrete light quanta that exhibit non-classical correlations–play a central role in quantum information and quantum communication technologies. It is a natural demand from technological applications on the intensity of the entangled photon pairs, such that sufficient signal strength can be achieved. Here we propose approaches based on klystron tubes that could potentially achieve stable amplification of the entangled photon pairs generated by spontaneous parametric down conversion (SPDC) and entanglement transfer.

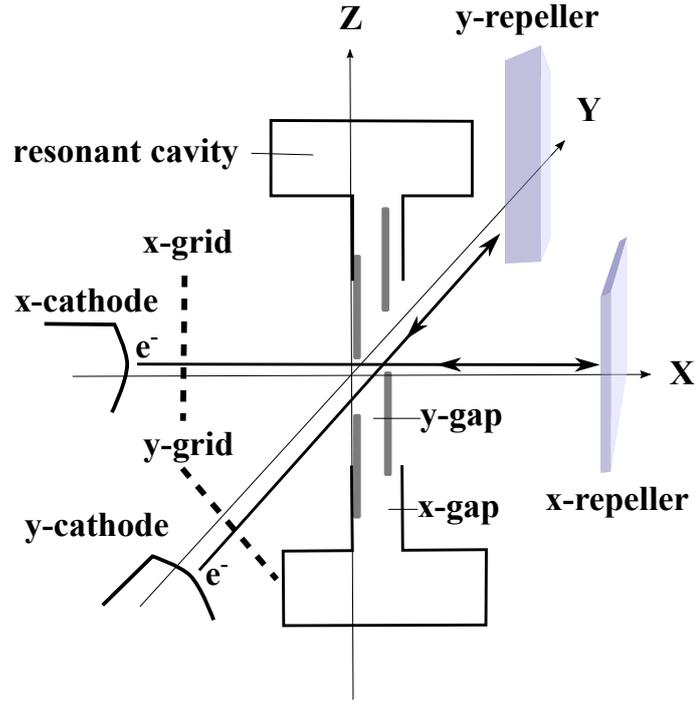

FIG. 1. The sketch of the proposed reflex klystron with two gaps in the *x* and *y* direction, which can be used to amplify entangled photon states.

## I. INTRODUCTION

Same as for the classical communication systems, a key step in road maps for the quantum communication and quantum information processing is to achieve high power output of the entangled quantum states from localized to flying qubits. It is thus important for applications of general purpose, to amplify the entangled states generated by various approaches such as by stimulated emission of entangled photons [1].

The klystron, which is based the velocity modulation of electron beams, is a widely applied electron tube for the generation and amplification of light of microwave frequency [2], and could be extended to operate at terahertz and optical frequencies [3–7]. Borrowing the traditional wisdom from the klystron amplifier, we propose approaches to amplify entangled states of photons.

## II. AMPLIFICATION OF ENTANGLED STATES BY DOUBLE-GAP KLYSTRON

Consider an input photon pair that is to be amplified is in the entangled state

$$|\Phi^{\pm}\rangle = \frac{1}{\sqrt{2}}\left(|X\rangle_1|Y\rangle_2 \pm |Y\rangle_1|X\rangle_2\right). \tag{1}$$

Here $|X\rangle$ or $|Y\rangle$ indicates a horizontally or vertically polarized photon, respectively. A reflex klystron as shown in Fig. 1 is proposed to amplify this entangled state. We show its feasibility through a mathematically unsophisticated quantum treatment of klystron amplifier [5–8]. Assume the initial state at the input port in the Fock representation as

$$|I\rangle = \frac{1}{\sqrt{2}}\left(|n0\rangle_1|0n\rangle_2 + |0n\rangle_1|n0\rangle_2\right)|\psi_x\rangle|\psi_y\rangle, \tag{2}$$

and the final state at the output port as

$$|F\rangle = \frac{1}{\sqrt{2}}\left(|n+1,0\rangle_1|0,n+1\rangle_2 + |0,n+1\rangle_1|n+1,0\rangle_2\right)|\psi'_x\rangle|\psi'_y\rangle, \tag{3}$$

where $|\psi_i\rangle$ and $|\psi'_i\rangle$, $i = x, y$ are the states of plane wave electrons before and after the interaction with photons in the gap region through spontaneous and stimulated processes. Using canonical quantization, we have the interaction Hamiltonian of the klystron in the velocity gauge

$$\hat{H}_{\text{int}} = -\frac{e}{2m}\int d\vec{r}\,\psi^{\dagger}(\vec{r})(\vec{p}\cdot\vec{A} + \vec{A}\cdot\vec{p})\psi(\vec{r}), \tag{4}$$

with the vector potential

$$\vec{A}(\vec{r}) = \Theta(\vec{r})\frac{i}{\omega d}\sqrt{\frac{\hbar\omega}{CV}}\left(\hat{a}_{\vec{k}}\vec{\varepsilon}_{\vec{k}}e^{i\vec{k}\cdot\vec{r}} + \hat{a}^{\dagger}_{\vec{k}}\vec{\varepsilon}^{*}_{\vec{k}}e^{-i\vec{k}\cdot\vec{r}}\right). \tag{5}$$

Here $d$ is the distance between the plates of the gap, $C = \varepsilon_0 A/d$ is the capacitance of the plates with effective area $A$, $\omega$ is the photon frequency. $e$, $m$ are the charge and mass of electron. $V$ is the quantization volume. $\Theta(\vec{r})$ is the rectangular function that is 1 for $0 \leq x, y \leq d$ inside the gap and 0 otherwise. We take the plane wave function for the electron beam and model the repeller as an infinitely high potential wall. The rate of spontaneous emission and stimulated emission and absorption can be calculated by the Fermi's golden rule as

$$\left\langle\frac{dP_{\text{em/ab}}}{dt}\right\rangle = \frac{2\pi}{\hbar}\left|\sum_M \frac{\langle F|\hat{H}_{\text{int}}|M\rangle\langle M|\hat{H}_{\text{int}}|I\rangle}{E_I - E_M \mp \hbar\omega + i\varepsilon}\right|^2 \delta(E_I - E_F \mp 2\hbar\omega), \tag{6}$$

where $E_I, E_M, E_F$ are the electron energies of the initial, intermediate and final states, respectively. Taking Eq. 6 and assume the input electron number $N_x = N_y = N$, the rate of net output

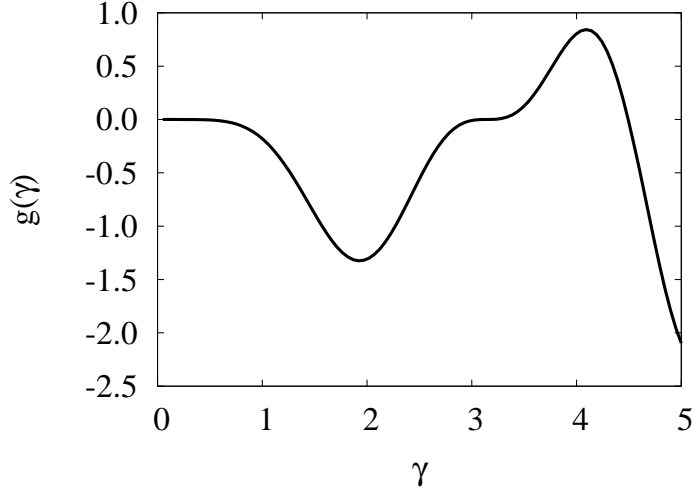

FIG. 2. Gain factor $g(\gamma)$ of the reflex klystron as a function of gap transit angle $\gamma$.

of photon pair number is calculated in the Appendix A to be

$$\left\langle \frac{dn_{\text{out}}}{dt} \right\rangle = N^2 \sum_{\vec{k}} \left( \left\langle \frac{dP_{\text{em}}}{dt} \right\rangle - \left\langle \frac{dP_{\text{ab}}}{dt} \right\rangle \right)$$
$$= \frac{32\pi^3 J_0^2}{m\hbar^4 C^2 d^4 \omega^7} \left[ 2E_p^2 + (\hbar\omega)^2 \right] (\gamma \cot\gamma - 1) \sin^4\gamma + \frac{16\pi^3 J_0^2 v^2}{\hbar^3 C^2 d^4 \omega^6}$$
$$= \frac{32\pi^3 J_0^2}{m\hbar^4 C^2 d^4 \omega^7} \left[ 2E_p^2 + (\hbar\omega)^2 \right] g(\gamma) + \frac{16\pi^3 J_0^2 v^2}{\hbar^3 C^2 d^4 \omega^6}. \tag{7}$$

Here $J_0 = \frac{evN}{V}$ is the electron flux, $E_p = n\hbar\omega$ is the total photon energy stored inside the resonant cavity. $\gamma = \frac{\omega d}{2v} = \frac{\theta_g}{2}$ is defined as the gap transit angle for the electron beam with initial velocity $v_x = v_y = v$. As shown in Fig. 2, the gain factor $g(\gamma)$ in Eq. 7 is a function of the gap transit angle, and peaks for $\gamma = 4.09$ with $g(4.09) = 0.843$. Suppose the input photon pair has degenerate frequency of 1 GHz, the width of the gap is 1 mm, the required velocity to reach $\gamma = 4.09$ is calculated as $v = 2.56 \times 10^{-4} c \ll c$, and the corresponding acceleration voltage is $U_0 = 1.72$ V, which is a moderate requirement.

Because the photons are measured independently of the electrons, we trace out the electronic degrees of freedom to obtain the reduced density matrix of the photons by themselves [9]

$$\begin{aligned}\rho_{\text{photon}} &= \text{Tr}_{\text{electron}}|F\rangle\langle F| \\ &= \frac{1}{2}(|n+1,0\rangle_1|0,n+1\rangle_2\langle n+1,0|_1\langle 0,n+1|_2 \\ &+ |0,n+1\rangle_1|n+1,0\rangle_2\langle 0,n+1|_1\langle n+1,0|_2 \\ &+ |n+1,0\rangle_1|0,n+1\rangle_2\langle 0,n+1|_1\langle n+1,0|_2 \\ &+ |0,n+1\rangle_1|n+1,0\rangle_2\langle n+1,0|_1\langle 0,n+1|_2)\,.\end{aligned} \qquad (8)$$

The degree of entanglement (concurrence) [9, 10] of this state is

$$\mathscr{C} = 1\,, \qquad (9)$$

such that the entanglement is maintained after the amplification. The reason relies on the fact that in the photon emission process

$$\begin{aligned}|n,0\rangle|0,n\rangle|\psi_x\rangle|\psi_y\rangle &\mapsto |n+1,0\rangle|0,n+1\rangle|\psi'_x\rangle|\psi'_y\rangle\,, \\ |0,n\rangle|n,0\rangle|\psi_x\rangle|\psi_y\rangle &\mapsto |0,n+1\rangle|n+1,0\rangle|\psi'_x\rangle|\psi'_y\rangle\,,,\end{aligned} \qquad (10)$$

the actions of the photon states $|n,0\rangle|0,n\rangle$ and $|0,n\rangle|n,0\rangle$ on the electron state $|\psi'_x\rangle|\psi'_y\rangle$ inside the gap area are indistinguishable, thus the entanglement is not broken during the amplification. The same processes take place for the absorption. However, we could see that for the other two Bell states

$$|\Psi^{\pm}\rangle = \frac{1}{\sqrt{2}}(|X\rangle_1|X\rangle_2 \pm |Y\rangle_1|Y\rangle_2)\,, \qquad (11)$$

the concurrence after amplification vanishes, which requires post-processing as proposed in the previous work [9].

As shown in Fig. 1, the proposed double gap reflex klystron requires a resonant cavity that is different to the conventional reflex klystron, because the double gap prohibits the cylindrically symmetric geometry of the klystron system. The space charge effect of the two electron beams along the $X$ and $Y$ directions could impose challenges to its implementation, though it is possible to avoid the two beams to intersect with each other by spatially separating them along the $Z$ direction. The more detailed simulation will be carried out in the follow-up work.

## III. AMPLIFICATION OF ENTANGLED STATES THROUGH PHOTON-ENTANGLED ELECTRON PAIR INTERACTION

To avoid the restriction of the cavity geometry and space charge effect of the double gap reflex klystron, it would be advantageous to consider the entanglement transfer [9] from the electron pair to the photon pair using conventional klystron and an entangled electron pair source. It has been demonstrated that entangled electron pairs can be generated through atomic and molecular processes, such as the photoelectron pair $(e_i, e'_i)$ from double photoionization (DPI) [11] and photoelectron-Auger-electron pair $(e_i, e_a)$ from photoionization-Auger decay (PAD) [12, 13]. For the photoionization-Auger decay process

$$A(S_I) \to A^+(S_M) + e_i \to A^{2+}(S_F) + e_a \tag{12}$$

during which an atom $A$ of in state of spin $S_I$ is photoionized to $A^+$ in ionic state of spin $S_M$, and then Auger decays to the dication $A^{2+}$ in state of spin $S_F$. It is found that the produced photoelectron-Auger-electron pair $(e_i, e_a)$ can form an entangled Werner state with the density matrix [12, 14]

$$\begin{aligned}\rho(\hat{u}_i, \hat{u}_a) &= p\sigma_1(\hat{u}_i, \hat{u}_a) + (1-p)\sigma_2, \\ 0 &\leq p \leq 1 ,\end{aligned} \tag{13}$$

where $\hat{u}_i$ and $\hat{u}_a$ are the spin quantization axis of the electron pair $(e_i, e_a)$,

$$\begin{aligned}\sigma_1 &= |\text{Bell}\rangle\langle\text{Bell}|, \\ \sigma_0 &= \frac{1}{4}\mathbf{I}_{4\times 4},\end{aligned} \tag{14}$$

and $p$ characterizes the degree of entanglement. It was shown [12] that for $S_I = S_F$ and $S_M > 0$,

$$p(S_I, S_M, S_F) = \frac{[3/4 - S_M(S_M+1) - S_I(S_I+1)]^2}{3S_M(S_M+1)}, \tag{15}$$

e.g. a singlet to singlet photoionization-Auger process can have $p(S_I = 0, S_M = 1/2, S_F = 0) = 1$, which corresponds to a maximally entangled state of flying spin qubit. For the photoelectron-Auger-electron pair with spin entanglement, it can be possible to transfer the entanglement to the photons by spontaneous and stimulated emission. Using the standard procedure via Foldy-Wouthuysen transformation, we have the nonrelativistic expansion of the Dirac equation [15,

16], which gives the Hamiltonian

$$\hat{H} = \frac{1}{2m}\left(-i\hbar\vec{\nabla} - e\vec{A}\right)^2 - \frac{e\hbar}{2m}\vec{\sigma}\cdot\vec{B} + e\phi - \frac{1}{8m^3c^2}\left(-i\hbar\vec{\nabla} - e\vec{A}\right)^4$$
$$- \frac{e^2\hbar^2}{8m^3c^4}\left(c^2B^2 - E^2\right) + \frac{e\hbar}{8m^3c^2}\left\{\vec{\sigma}\cdot\vec{B}, (-i\hbar\vec{\nabla} - e\vec{A})^2\right\}$$
$$- \frac{e\hbar}{4m^2c^2}\vec{\sigma}\cdot[\vec{E}\times(-i\hbar\vec{\nabla} - e\vec{A})] - \frac{e\hbar^2}{8m^2c^2}\vec{\nabla}\cdot\vec{E}. \quad (16)$$

The second order interaction from the Hamiltonian term $\frac{e\hbar}{2m}\vec{\sigma}\cdot\vec{B}$ could transfer the spin entanglement of electrons to the polarization entanglement of photons to the leading order, the calculation will be presented in the follow-up study.

## IV. CONCLUSION

We proposed an approach to amplify the entangled states by klystron through velocity modulated electron beams. We also point out the limitations of the proposed double gap klystron amplifier that it needs post-processing for the amplification of two of the four Bell states.

### ACKNOWLEDGMENTS

Z.L. is thankful to the Volkswagen Foundation for a Paul-Ewald postdoctoral fellowship, and thanks Matthew Ware, Andreas Kaldun for stimulating discussions.

### Appendix A: Amplification by a quantized 2D reflex klystron

We treat the reflex klystron illustrated in Fig. 1 as a 2D resonant cavity-capacitor system by describing the gap as capacitor, which has an eigen frequency $\omega$ of oscillation. The voltage of the capacitors in parallel connection satisfies the equation of motion

$$\sum_{i=x,y}\ddot{U}_i + \omega^2 U_i = 0, \quad (A1)$$

which can be derived by the Hamiltonian

$$H = \frac{1}{2}\sum_i \frac{C_i}{\omega^2}\dot{U}_i^2 + C_i U_i^2 \quad (A2)$$

through Hamiltonian canonical equations. Define canonical coordinate $Q_i = U_i$, its conjugate momentum is thus $P_i = \frac{\partial \mathscr{L}}{\partial \dot{Q}_i} = \frac{C}{\omega^2}\dot{U}_i$, and their quantized operator satisfy the Dirac canonical quantization condition $[\hat{Q}_i, \hat{P}_j] = i\hbar \delta_{ij}$. The Hamiltonian under canonical quantization is thus

$$\hat{H} = \frac{1}{2}\sum_i \frac{\omega^2}{C_i}\hat{P}_i^2 + C_i Q_i^2 \tag{A3}$$

And by defining the creation and annihilation operator of the light field through standard field quantization procedure, we obtain the voltage operator expressed in terms of light field operators,

$$\hat{U}(\vec{r}) = \sqrt{\frac{\pi \hbar \omega}{CV}} \sum_{\vec{k},\vec{\varepsilon}} \left( \hat{a}_{\vec{k}} \vec{\varepsilon}_{\vec{k}} e^{i\vec{k}\cdot\vec{r}} + \hat{a}_{\vec{k}}^\dagger \vec{\varepsilon}_{\vec{k}}^* e^{-i\vec{k}\cdot\vec{r}} \right), \tag{A4}$$

where $V = L^2$ is the quantization volume of the 2D system. The interaction energy between the gap and the electron is

$$E_{\text{int}} = -e\vec{E}\cdot\vec{r} = -e\sum_i \frac{U_i}{d} r_i, \tag{A5}$$

and the light field-electron interaction Hamiltonian is obtained in the length gauge as

$$\hat{H}_{\text{int}} = -\frac{e}{d}\sum_{\vec{k},\vec{\varepsilon}} \sqrt{\frac{\pi\hbar\omega}{CV}} \int d\vec{r}\, \hat{\psi}^\dagger(\vec{r})\vec{r}\cdot \left( \hat{a}_{\vec{k}} \vec{\varepsilon}_{\vec{k}} e^{i\vec{k}\cdot\vec{r}} + \hat{a}_{\vec{k}}^\dagger \vec{\varepsilon}_{\vec{k}}^* e^{-i\vec{k}\cdot\vec{r}} \right) \hat{\psi}(\vec{r}), \tag{A6}$$

which is equivalent to the Hamiltonian in Eq. 4 under velocity gauge by employing the Heisenberg equation $\frac{i\hbar}{m}\hat{p} = [\hat{r}, \hat{H}_0]$.

Now we calculate the emission and absorption rate $\left\langle \frac{dP_{\text{em/ab}}}{dt} \right\rangle$. For the photon emission process, we consider the intermediate states as the state

$$|M\rangle = \frac{1}{\sqrt{2}} \left( |n+1,0\rangle_1 |0,n\rangle_2 + |0,n\rangle_1 |n+1,0\rangle_2 \right) |\psi'_x\rangle |\psi_y\rangle, \tag{A7}$$

and the state with $x, y$ interchanged. From Fermi's golden rule, we have for the emission process

$$\left\langle \frac{dP_{\text{em}}}{dt} \right\rangle = \frac{2\pi}{\hbar} \left| \sum_M \frac{\langle F|\hat{H}_{\text{int}}|M\rangle \langle M|\hat{H}_{\text{int}}|I\rangle}{E_I - E_M - \hbar\omega + i\varepsilon} \right|^2 \delta(E_I - E_F - 2\hbar\omega)$$

$$= 2\frac{2\pi}{\hbar}(n+1)^2 \left| \sum_{k'_x} \left(\frac{e}{Ld}\right)^2 \frac{\pi\hbar\omega}{C} \frac{\langle \psi'_x|x|\psi_x\rangle \langle \psi'_y|y|\psi_y\rangle}{E_I - E_M - \hbar\omega + i\varepsilon} \right|^2 \delta(E_I - E_F - 2\hbar\omega). \tag{A8}$$

Here $|\psi_i\rangle$ and $|\psi'_i\rangle$ are the electron state before and after emission. We assume the electron wave function to be

$$\psi_x(x) = \frac{1}{\sqrt{2L}} (e^{ik_x x} + e^{-ik_x x}), \tag{A9}$$

with $k_x = \frac{2v_x\pi}{L}$. The repeller apart from the gap by a drifting distance $l$ is treated as an infinitely high potential at $x = l$, it sets the boundary condition for the electron wave function, such that $k_x = \frac{v_x\pi}{2l}$. Employing the rotating wave approximation, the matrix element $\langle \psi'_x | x | \psi_x \rangle$ can be calculated as $\langle \psi'_x | x | \psi_x \rangle = \int_{-L/2}^{0} + \int_{0}^{d} + \int_{d}^{l} + \int_{l}^{L/2} [dx \psi'_x(x)^* x \psi_x(x)]$, and it follows

$$\left| \langle \psi'_x | x | \psi_x \rangle \right|^2 = \frac{2}{L^2} \frac{\sin^2[(k-k')\frac{d}{2}]}{(k-k')^4}. \tag{A10}$$

The matrix element is the same for the $y$ component. Using the relation $\sum_{k'} = \frac{L}{2\pi} \int dk'$ and contour integral, Eq. A8 can be calculated as

$$\left\langle \frac{dP_{\text{em}}}{dt} \right\rangle = \frac{16\pi^3 m^2 e^2 \omega^2 (n+1)^2}{L^6 \hbar^3 C^2 d^4} \frac{\sin^4[(k-k')\frac{d}{2}]}{k'^2(k-k')^8} \delta(E_I - E_F - 2\hbar\omega). \tag{A11}$$

Here we assume $k_x = k_y = k$ and $k'_x = k'_y = k'$. Employing the condition of momentum and energy conservation

$$\begin{aligned} k'_i &= k_i - \Delta k_i \\ E'_i &= E_i - \hbar\omega, \end{aligned} \tag{A12}$$

we obtain the relation

$$k'_i = k_i[1 - \alpha(1 + \frac{1}{2}\alpha)], \tag{A13}$$

with $\alpha = \frac{m\omega}{\hbar k^2} = \frac{\hbar\omega}{mv^2} \ll 1$. Expanding terms in Eq. A11 with respect to $\alpha$, we can obtain the rate for the emission process

$$\left\langle \frac{dP_{\text{em}}}{dt} \right\rangle = \frac{16\pi^3 e^2 \hbar^5 k^6}{L^6 C^2 d^4 m^6 \omega^6} (n+1)^2 [1 - 2\alpha(1 - \gamma \cot \gamma)] \sin^4 \gamma \delta(E_I - E_F - 2\hbar\omega), \tag{A14}$$

and similarly for the absorption process

$$\left\langle \frac{dP_{\text{ab}}}{dt} \right\rangle = \frac{16\pi^3 e^2 \hbar^5 k^6}{L^6 C^2 d^4 m^6 \omega^6} n^2 [1 + 2\alpha(1 - \gamma \cot \gamma)] \sin^4 \gamma \delta(E_I - E_F - 2\hbar\omega), \tag{A15}$$

where $\gamma = \frac{kd\alpha}{2} = \frac{\omega d}{2v}$ is the gap transit angle of the electron. The output rate of entangled photon pair is thus $\left\langle \frac{dP_{\text{out}}}{dt} \right\rangle = \left\langle \frac{dP_{\text{em}}}{dt} \right\rangle - \left\langle \frac{dP_{\text{ab}}}{dt} \right\rangle$. The output number rate of entangled photon pair is $\left\langle \frac{dn_{\text{out}}}{dt} \right\rangle = N^2 \sum_{\vec{k}} \left\langle \frac{dP_{\text{out}}}{dt} \right\rangle$, where we denote the input electron number as $N_x = N_y = N$. The leading contribution of output rate of entangled photon pair number is thus obtained for the stimulated process, i.e. the process through which the electrons gain energy from the field in the cavity or lose energy to the cavity in the presence of light field in the cavity, as

$$\left\langle \frac{dn_{\text{out}}}{dt} \right\rangle_{\text{st}} = \frac{64\pi^3 J_0^2 E_p^2}{m\hbar^4 C^2 d^4 \omega^7} g(\gamma), \tag{A16}$$

which depends on the stored photon energy $E_p$ in the cavity. $g(\gamma) \equiv (\gamma \cot \gamma - 1)\sin^4 \gamma$ is the gain factor. The output number rate for the spontaneous process, i.e. the process takes place without presence of initial light field in the cavity, is

$$\left\langle \frac{dn_{\text{out}}}{dt} \right\rangle_{\text{sp}} = \frac{32\pi^3 J_0^2 (\hbar\omega)^2}{m\hbar^4 C^2 d^4 \omega^7} g(\gamma) + \frac{16\pi^3 J_0^2 v^2}{\hbar^3 C^2 d^4 \omega^6}. \tag{A17}$$

---